\documentclass[twoside,slac_one]{revtex4}
\usepackage{graphicx}
\usepackage{epsf} 
\usepackage{fancyhdr}
\usepackage{amsmath} % American Mathematics Society standards
\usepackage{bm}% bold math
\usepackage{amsxtra}
\usepackage{amssymb}
\usepackage{amsthm}
\usepackage{latexsym}
\usepackage{lscape}

\newcommand{\BDK}{\ensuremath{B^- \to D K^-}}

\begin{document}

%Title of paper
\title{First ADS Analysis of $\mathbf{B^- \to D^0 K^-}$ Decays in Hadron Collisions}

% Repeat the \author .. \affiliation  etc. as needed
%
% \affiliation command applies to all authors since the last
% \affiliation command. The \affiliation command should follow the
% other information

\author{Paola Garosi, for the CDF Collaboration}
\affiliation{University of Siena and INFN of Pisa, Italy}

\begin{abstract}
We report the first measurement of branching fractions and $CP$-violating asymmetries of suppressed \BDK decays in hadron collisions, using the approach proposed by Atwood, Dunietz, and Soni (ADS) to determine the CKM angle $\gamma$ in 7.0 fb$^{-1}$ of data. The ADS parameters are determined with accuracy comparable with $B$-factory measurements and significantly improve the global knowledge of the angle $\gamma$.
\end{abstract}

%\maketitle must follow title, authors, abstract
\maketitle

\thispagestyle{fancy}

% body of paper here - Use proper section commands
% References should be done using the \cite, \ref, and \label commands
% Put \label in argument of \section for cross-referencing
%\section{\label{}}

\section{Introduction}
The {\it Cabibbo-Kobayashi-Maskawa matrix} contains the couplings of the weak interaction among the quarks. Knowledge of its parameters allows both to test the Standard Model and to probe New Physics scenarios. 
The unitarity condition of the CKM matrix~\cite{ref:CKM} allows to create the so called ``Unitarity Triangle", whose angles are $\alpha$, $\beta$ and $\gamma$.
While $\alpha$ and $\beta$ have been determined to a good level of precision~\cite{ref:hfag}, the measurement of $\gamma= arg (- V_{ud} V^*_{ub} / V_{cd} V^*_{cb})$ is still limited by the smallness of the branching ratios involved in the processes used to measure it, and its relative uncertainty varies between $15$ and  $20\%$, depending on the method used to combine the experimental results~\cite{ref:hfag,ref:utfit,ref:ckmfitter}.

There are several methods to measure $\gamma$, the ones with the smallest theoretical uncertainties make use of the tree-level dominated $B^- \to D K^-$ decays (where $D$ labels either $D^0$ or $\overline{D}^0$ mesons) ~\cite{ref:glw1,ref:ads1,ref:ggsz}. The angle $\gamma$ appears as the relative weak phase between two amplitudes, the favored $b \to c \bar{u} s$ of the $B^- \to D^0 K^-$ (whose amplitude is proportional to the CKM elements $V_{cb} V_{us}$) and the color-suppressed $b \to u \bar{c} s$ of the $B^- \to \overline{D}^0 K^-$ (whose amplitude is proportional to $V_{ub} V_{cs}$). The interference between $D^0$ and $\overline{D}^0$ decaying into the same final state leads to measurable $CP$-violating effects.

According to the final state of the $D$ decay, the following methods have been suggested to infer $\gamma$:
\begin{itemize}
\item 
\textit{GLW (Gronau-London-Wyler) method}~\cite{ref:glw1,ref:glw2}, which uses $CP$ eigenstates of $D^0$, as $D^0_{CP^+} \rightarrow K^+ K^-, \pi^+ \pi^-$ and $D^0_{CP-} \rightarrow K^0_s \pi^0, K^0_s \phi, K^0_s \omega$;
\item 
\textit{ADS (Atwood-Dunietz-Soni) method}~\cite{ref:ads1,ref:ads2}, which uses the doubly Cabibbo suppressed mode $D^0 \rightarrow K^+ \pi^-$;
\item 
\textit{GGSZ (or Dalitz) method}~\cite{ref:ggsz,ref:ads2}, which uses three body decays of $D^0$, as 
$D^0 \rightarrow~K^0_s \pi^+ \pi^-$.
\end{itemize} 

All mentioned methods require no tagging or time-dependent 
measurements, and many of them only involve charged particles in 
the final state. They are therefore particularly well-suited to the hadron 
collider environment, where the large production of $B$ mesons can be exploited.
The use of a specialized trigger based on online detection of a secondary vertex (SVT trigger~\cite{ref:trigger}) allows the selection of pure $B$ meson samples.%, with a resolution of about 20 MeV$/$c$^2$.

We will describe in more details the ADS and GLW methods, for which CDF reports the first results in hadron collisions.

\section{CDF II detector and trigger}
The CDF experiment is located at the Tevatron, a $\sqrt{s} = 1.96$ TeV $p\bar{p}$ collider. 
The detector~\cite{ref:cdf1} is a multipurpose magnetic spectrometer surrounded by calorimeters and muon detectors. 
Most relevant features for the measurement described here are the tracking, the particle-identification (PID) detectors and the trigger system. \\
The tracking system provides a determination of the decay point of particles with 15 $\mu$m resolution in the transverse plane using six layers of double-sided silicon-microstrip sensors at radii between 2.5 and 22 cm from the beam. 
A 96-layer drift chamber extending radially from 40 to 140 cm from the beam provides the reconstruction of three-dimensional charged-particles trajectories with excellent transverse momentum resolution, $\sigma_{p_T} / p^2_{T} = 0.1\% $ $1/GeV/c$. Specific ionization measurements in the chamber allow $1.5\sigma$ separation between charged kaons and pions, approximately constant at momenta larger than 2 GeV$/c$.

A three-level trigger system~\cite{ref:trigger} selects events enriched in decays of long-lived particles by exploiting the presence of displaced tracks in the event and measuring their impact parameter with 30 $\mu$m resolution. The trigger requires the presence of two charged particles with transverse momenta greater than 2 GeV$/c$, impact parameters greater than 100 microns and basic cuts on angular separation and scalar sum of momenta of the particles.

\section{The Atwood-Dunietz-Soni method}
In the ADS method~\cite{ref:ads1,ref:ads2} $\gamma$ appears in the interference between the
$B^- \to D^0 K^-$ (\textit{color favored}), followed by the $D^0 \to K^+ \pi^-$ (\textit{doubly Cabibbo suppressed}), and $B^- \to \overline{D}^0 K^-$ (\textit{color suppressed}), followed by the $ \overline{D}^0 \to K^+ \pi^-$ (\textit{Cabibbo favored}).
\begin{figure}[!ht]
\centering
\includegraphics[width=3.3in]{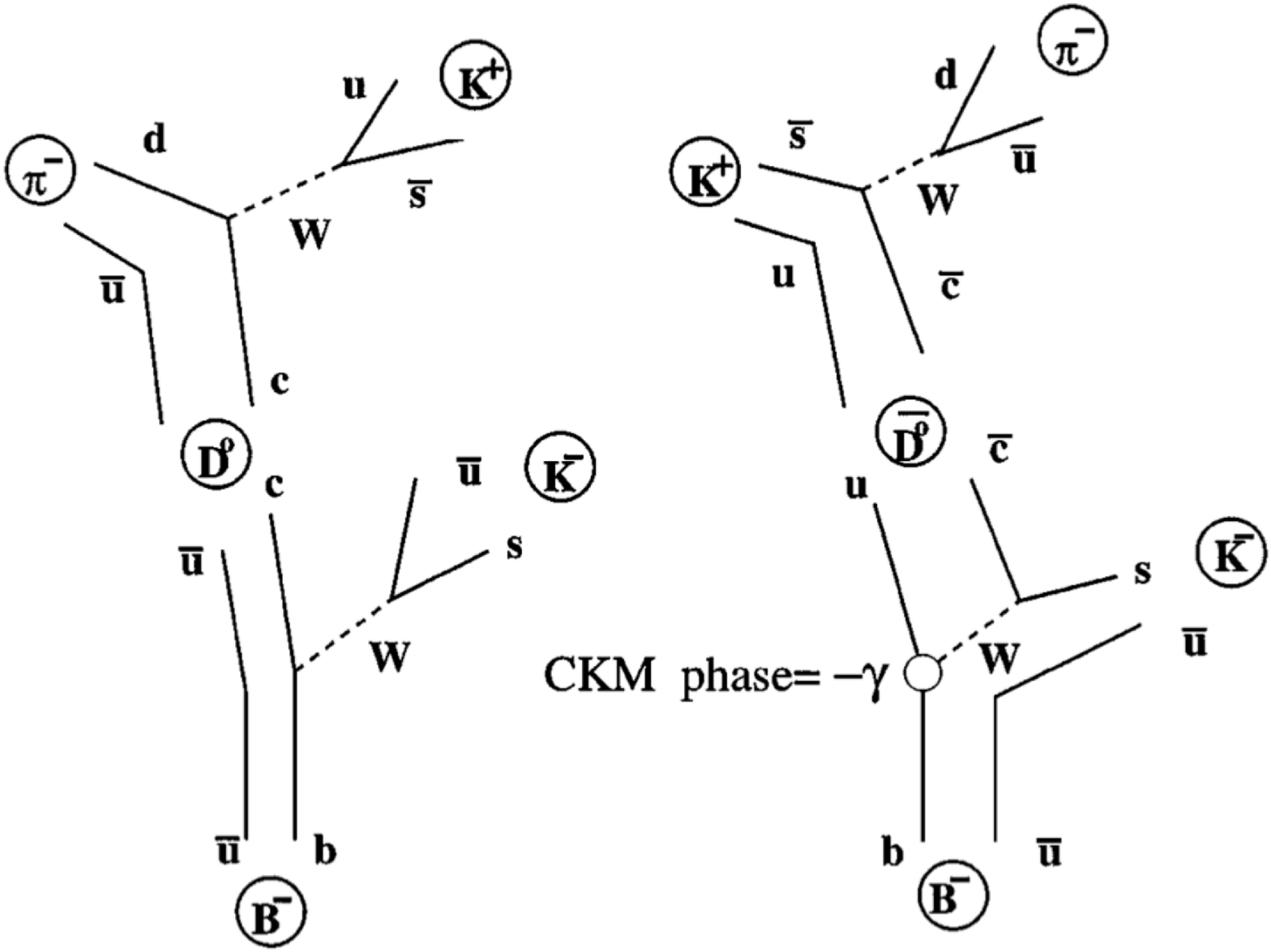}
\caption{Leading order of the diagram of the two interfering processes (from~\cite{ref:ads1}): $B^- \rightarrow D^0 K^-$ (color allowed), followed by $D^0 \rightarrow K^+\pi^-$ (doubly Cabibbo suppressed) and $B^- \rightarrow \overline{D}^0 K^-$ (color suppressed), followed by $\overline{D}^0 \rightarrow K^+\pi^-$ (Cabibbo allowed).}
\label{fig:after_cuts}
\end{figure}
Since $D^0$ and $\overline{D}^0$ are indistinguishable, we can only reconstruct the final state $[K^+ \pi^-]_D K^-$ and measure the direct $CP$- violating asymmetry. For simplicity we call ``suppressed" ({\it sup}) this final state.
Because the interfering amplitudes are of the same order of magnitude, we expect large $CP$-violating effects:
$$
\left| \frac{\mathcal{M} (B^-\to \overline{D}^0(\to K^+ \pi^-) K^-)}{\mathcal{M} (B^-\to D^0(\to K^+ \pi^-) K^-)}\right|^2 
\sim \left| \frac{V_{ub} V^*_{cs}}{V_{cb}V^*_{us}}\right| ^2 |f_{\it col\_sup}|^2 
\frac{\mathcal{B} (\overline{D}^0\to K^+ \pi^-) } {\mathcal{B} (D^0 \to K^+ \pi^-)} \sim 2,
$$
where $\mathcal{M}$ and $\mathcal{B}$ are the amplitude and the branching ratio of the decay, and $f_{\it col\_sup}$ is the color suppression factor $\sim 1/3$.

The direct $CP$-asymmetry,
$$
\displaystyle A_{ADS}  = \frac{\mathcal{B}(B^-\rightarrow [K^+\pi^-]_{D}K^-)-\mathcal{B}(B^+\rightarrow [K^-\pi^+]_{D}K^+)}{\mathcal{B}(B^-\rightarrow [K^+\pi^-]_{D}K^-)+\mathcal{B}(B^+\rightarrow [K^-\pi^+]_{D}K^+)},
$$ 
can be written in terms of the decay amplitudes and phases 
$$ A_{ADS}  =  \frac{2r_B r_D\sin{\gamma}\sin{(\delta_B+\delta_D)}}{r_D^2 + r_B^2 + 2r_Dr_B \cos{\gamma}\cos{(\delta_B+\delta_D)}},$$
where $r_B = |\mathcal{M}(B^- \rightarrow \overline{D}^0 K^-)/\mathcal{M}(B^- \rightarrow {D}^0 K^-)|$, $\delta_B = arg[\mathcal{M}(B^- \rightarrow \overline{D}^0 K^-)/\mathcal{M}(B^- \rightarrow {D}^0 K^-)]$, $r_D= |\mathcal{M}({D}^0 \rightarrow K^+\pi^-)/\mathcal{M}(\overline{D}^0 \rightarrow K^+\pi^-)|$ and $\delta_D = arg[\mathcal{M}({D}^0 \rightarrow K^+\pi^-)/\mathcal{M}(\overline{D}^0 \rightarrow K^+\pi^-)]$.% are the corresponding amplitude ratio and strong phase difference of the $D$ meson.\\

The denominator corresponds to another physical observable, the ratio between suppressed and favored ({\it fav}) events, the latter coming from the decay channel
$B^- \to D^0 K^- $ (\textit{color favored}), followed by $D^0 \to K^- \pi^+$ (\textit{Cabibbo favored}): 
%$R_{ADS}  =  r_D^2 + r_B^2 + 2r_Dr_B \cos{\gamma}\cos{(\delta_B+\delta_D)} $
$$
R_{ADS}  =    \frac{\mathcal{B}(B^-\rightarrow [K^+ \pi^-]_{D}K^-)+\mathcal{B}(B^+\rightarrow [K^-\pi^+]_{D}K^+)}{\mathcal{B}(B^-\rightarrow [K^- \pi^+]_{D}K^-)+\mathcal{B}(B^+\rightarrow [K^+\pi^-]_{D}K^+)}. 
$$

The ratios of suppressed and favored decays for separated charges of $B$ can be measured as well, providing a set of statistically independent observables~\cite{ref:rpm}: 
$$
R^{\pm} = \frac{\mathcal{B}(B^{\pm} \rightarrow [K^{\mp} \pi^{\pm}]_{D} K^{\pm})}{\mathcal{B}(B^{\pm} \rightarrow [K^{\pm} \pi^{\mp}]_{D} K^{\pm})}.
$$
%
%As for the DCS label, we will use the label CF to identify the final state $[K^- \pi^+]_D K^-$ and the decay mode $B \to D^0_{CF} K$.\\
%
We can measure the corresponding quantities $A_{ADS}$, $R_{ADS}$ and $R^{\pm}$ also for the 
$B^- \to D \pi^-$ mode, for which sizeable asymmetries may be found~\cite{ref:hfag}. The maximum possible value of the asymmetry is $A_{\rm max} = 2r_B r_D/(r^2_B + r^2_D)$, where $r_B$ can be $r_B(K)$ or $r_B(\pi)$. Taking into account the CKM structure of the contributing processes, we expect that $r_B(\pi)$ is suppressed by a factor $|V_{cd}V_{us}/V_{ud}V_{cs}| \sim \tan^2 \theta_C$ with respect to $r_B(K)$, where $\theta_C$ is the Cabibbo angle, and we assume the same color suppression factor for both $DK$ and $D\pi$ modes.
Using $r_B(K) = 0.103 ^{+0.015}_{-0.024}$~\cite{ref:hfag}, $r_B(\pi) \sim 0.005$~\cite{ref:hfag}, and $r^2_D = (3.80 \pm 0.18)\times 10^{-3}$~\cite{ref:pdg}, we expect $A_{\rm max}(K) \approx 0.90 $ and $A_{\rm max}(\pi) \approx 0.16$.

\section{Sample selection and fit strategy}

\begin{figure}[!h]
%\begin{center}
\centering
\includegraphics[width=2.9in]{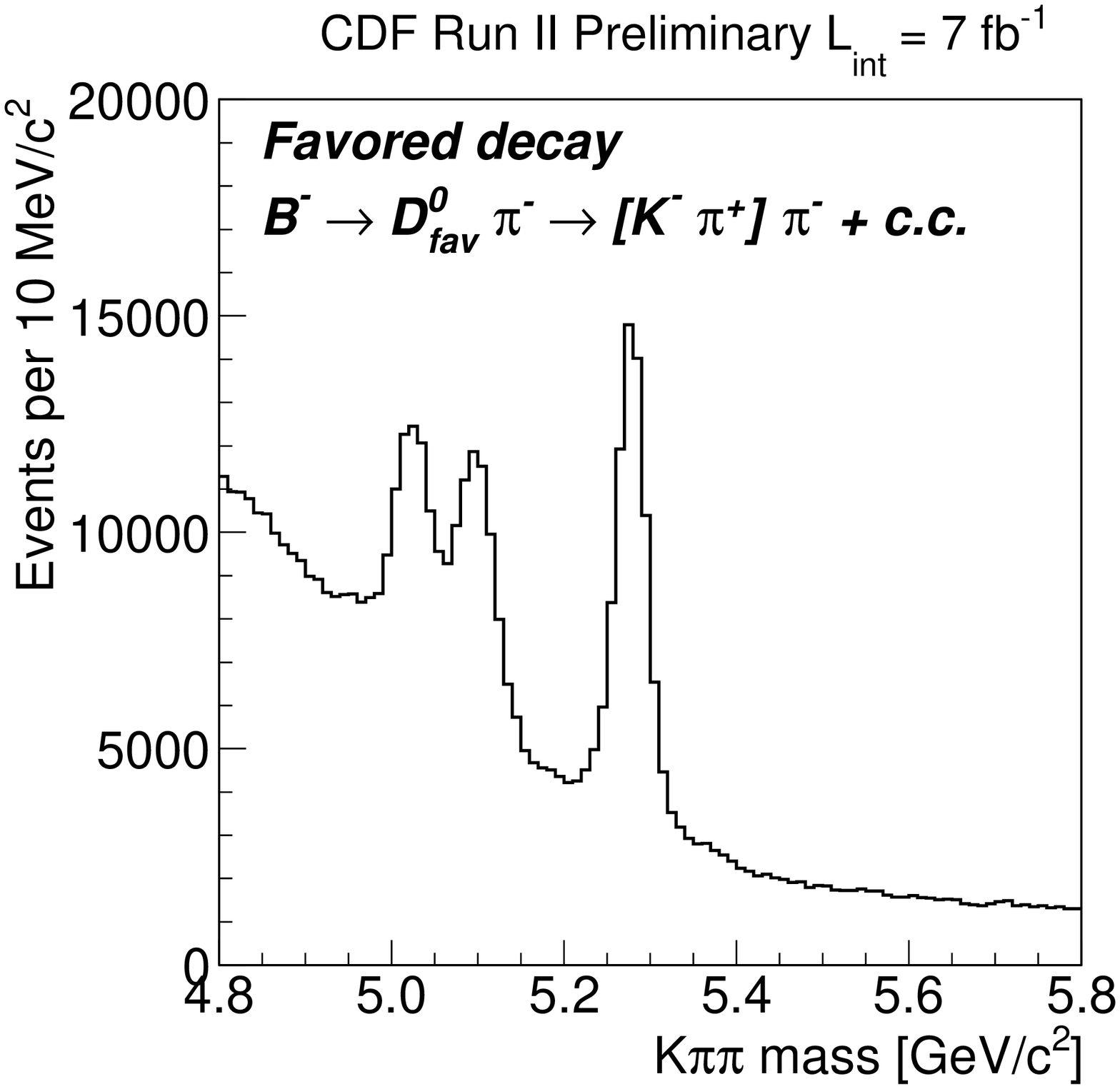} 
\hspace{0.5cm}    
\includegraphics[width=2.9in]{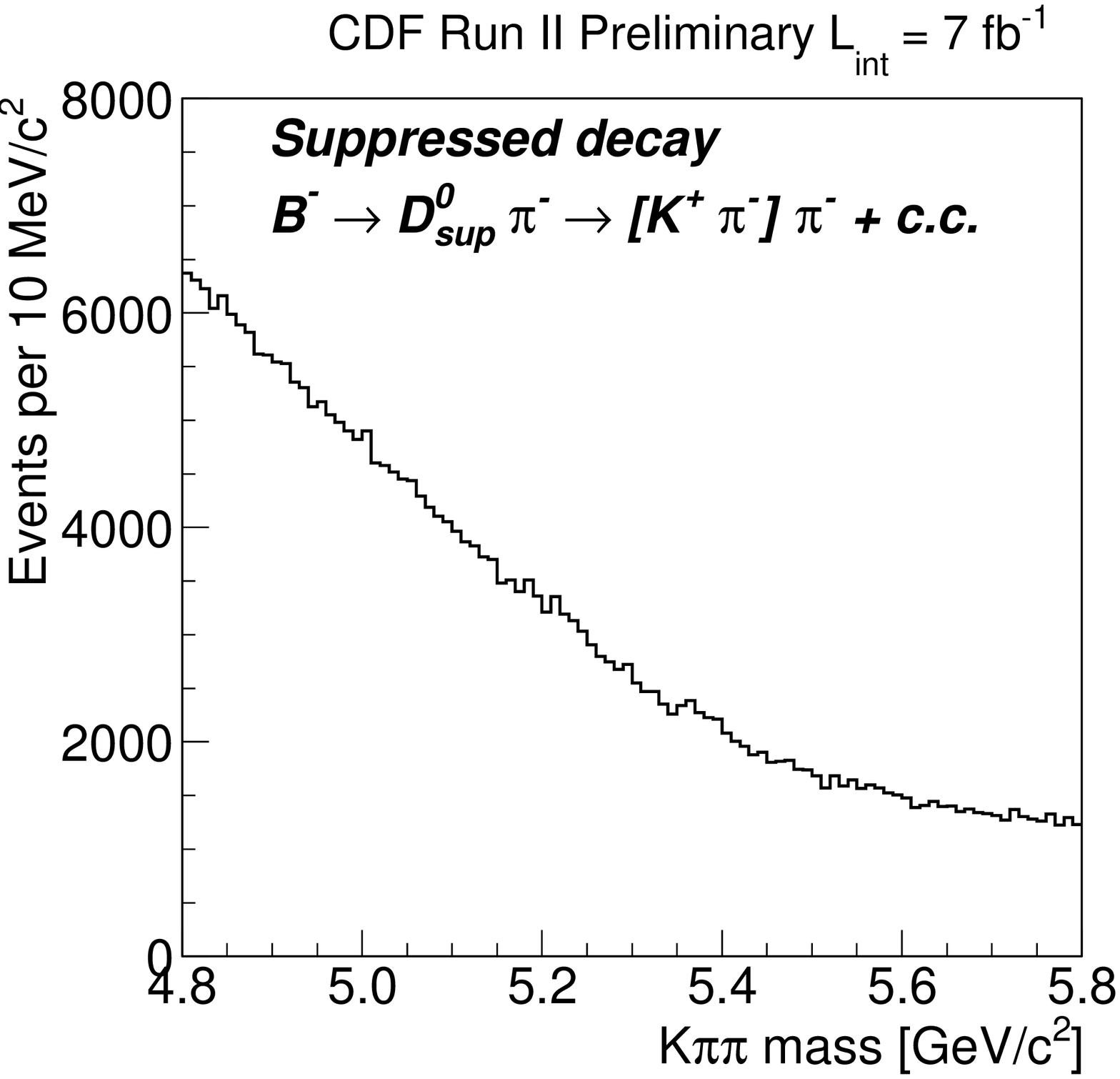}  
\caption{Mass distributions of $B^- \to D h^-$ ($h$ is $\pi$ or $K$) candidates, with a nominal pion mass assignment to the charged track from the $B$ meson decay, for each reconstructed
decay mode, favored on the left and suppressed on the right. } \label{fig:before_cuts}
%\end{center}
\end{figure}

The mass distributions of the favored and suppressed modes, using a sample of 7~fb$^{-1}$ of data, with a nominal pion mass assignment to the charged track from the $B$ meson decay, are shown in Fig.~\ref{fig:before_cuts}. 

A $B^- \to D \pi^-$ favored signal is
visible at the expected mass of about 5.279 GeV$/c^2$. 
Events from $B^- \to D K^-$ decays are expected to cluster in smaller and wider
structures, located about 50 MeV$/c^2$ below the $B^- \to D \pi^-$ signal.
The $B^- \to D \pi^-$ and $B^- \to D K^-$ suppressed signals appear to be buried in the combinatorial background. 
Suppression of the combinatorial background is obtained through an optimization of the selection requirements focused on finding a signal of the $B^- \to D_{\it sup} \pi^-$ mode. 
Since the $B^- \to D_{\it fav} \pi^-$ mode has the same topology of the suppressed one, but with larger statistics, the optimization uses signal ($S$) and background ($B$) directly from favored data to determine a selection that maximizes the figure of merit $S/(1.5+\sqrt{B})$~\cite{ref:punzi}. 

Several variables have been chosen to discriminate signal from background~\cite{ref:dcsPaper}, the most important being a threshold on the \textit{three-dimensional vertex quality} $\chi^2_{3D}$, which exploits the 3D silicon-tracking to resolve multiple vertices along the beam direction and to reject fake tracks, and the $B$ {\it isolation}. 
% It allows a background reduction by a factor of two and has small inefficiency on signal (less than 10\%). 
%The $B$ isolation corresponds to the fraction of momentum carried by the $B$ meson, which is usually greater than the momentum carried by lighter mesons.
Another important variable is the \textit{decay length of the $D$ with respect to the $B$}, which allows rejection of most of the $B^- \to hhh$ backgrounds, where $h$ is either the charged $\pi$ or $K$.  
All variables and threshold values applied are described in ref.~\cite{ref:dcsPaper}. 
The resulting invariant mass distributions of favored and suppressed modes are reported in Fig.~\ref{fig:after_cuts}, where the combinatorial background is significantly reduced in the $B^-$ mass region, allowing a signal structure to be seen.
\begin{figure}[!ht]
%\begin{center}
\centering
\includegraphics[width=2.9in]{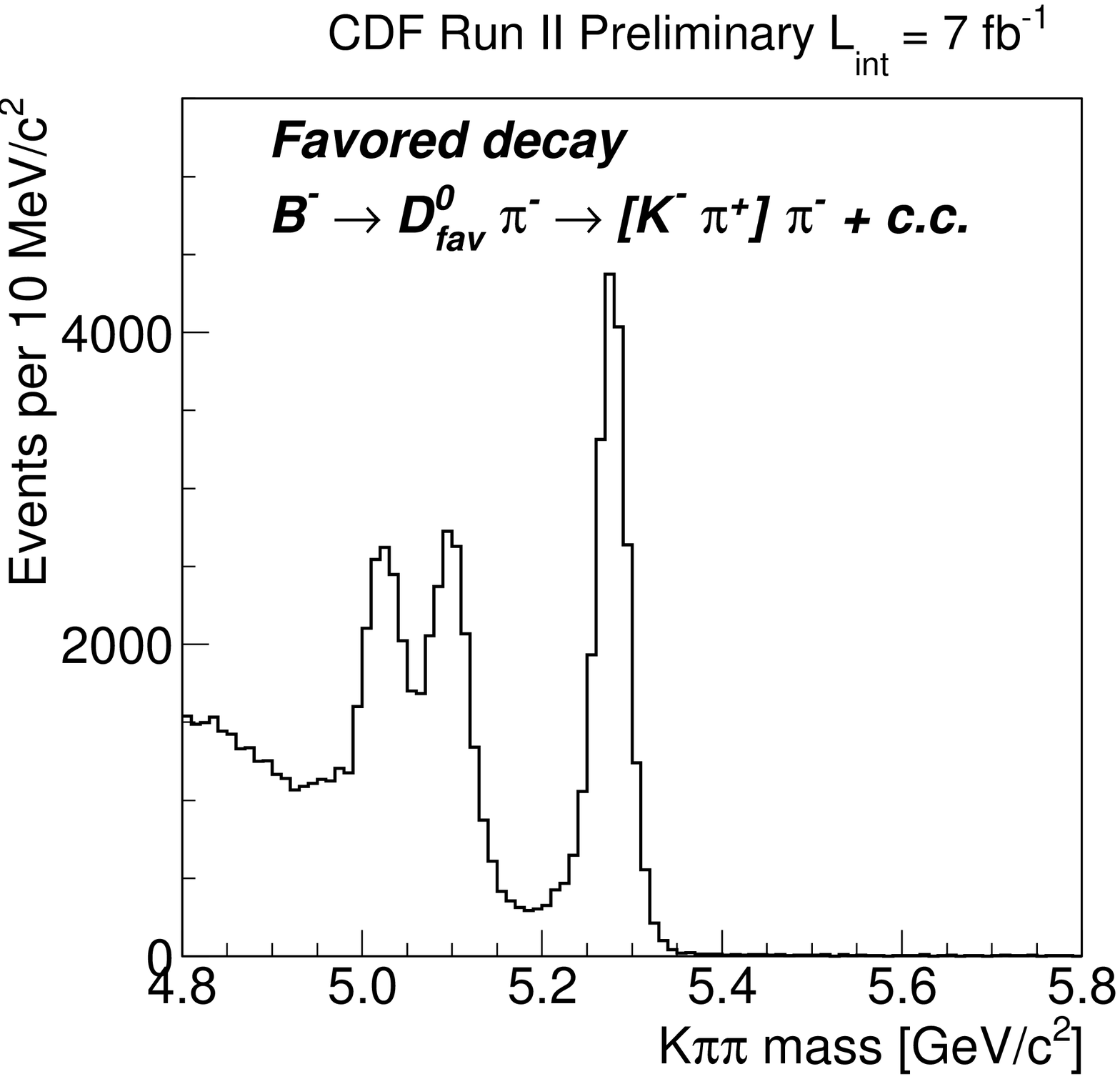}
\hspace{0.5cm}
\includegraphics[width=2.9in]{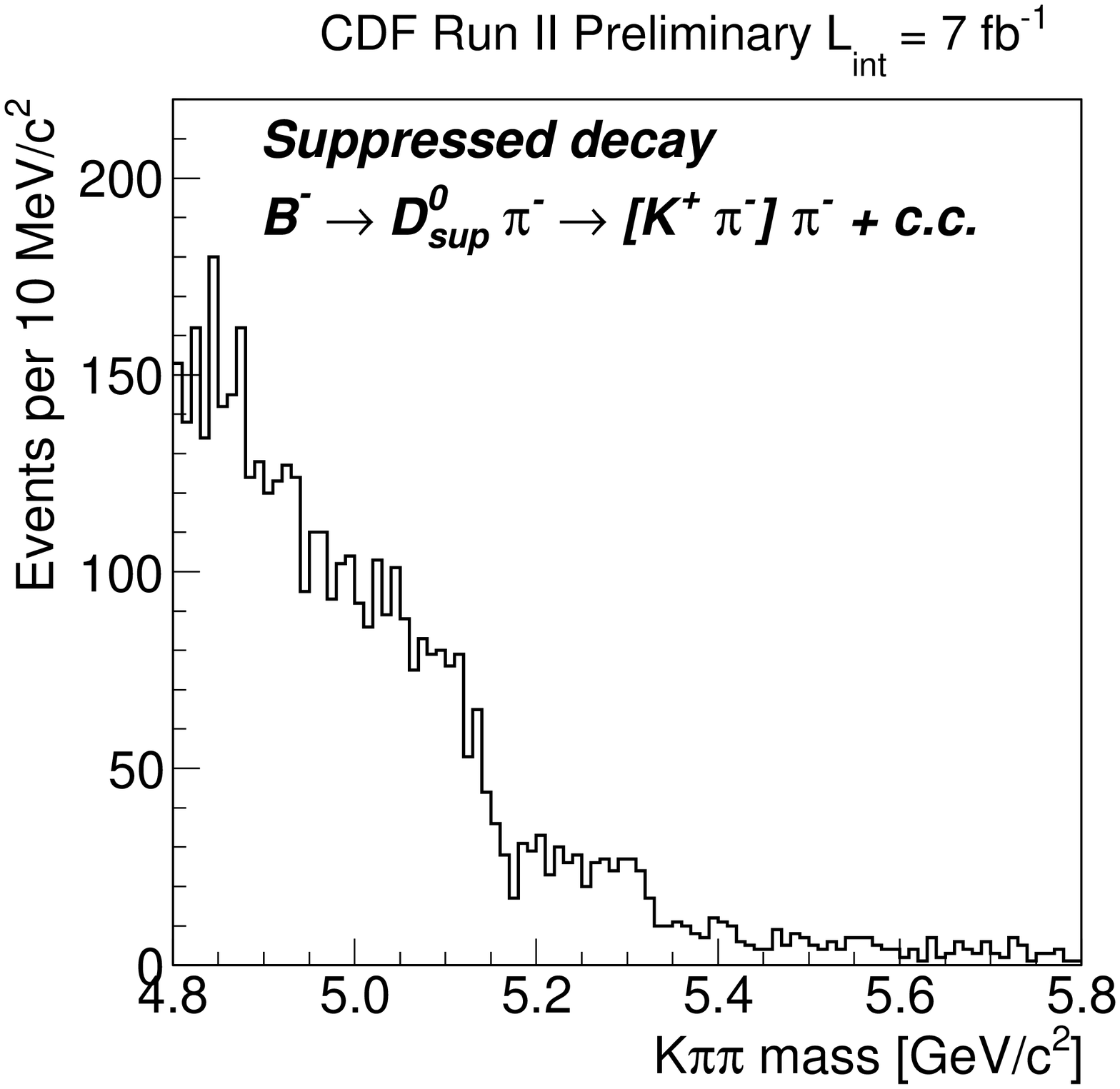} 
\caption{Mass distributions of $B^- \to D h^-$ candidates, with a nominal pion mass assignment to the charged track from the $B$ meson decay, for each reconstructed
decay mode, favored on the left and suppressed on the right, after the optimization of the selection.} \label{fig:after_cuts}
%\end{center}
\end{figure}
%\clearpage

An unbinned maximum likelihood fit, exploiting mass and particle identification information is simultaneously performed %~\cite{ref:pubnote} 
on both favored and suppressed samples, to separate the $B^- \to D K^-$ contributions from the $B^- \to D \pi^-$ signals and the combinatorial and physics backgrounds. 
%The particle identification information is provided by the specific ionization (dE$/$dx) of the CDF drift chamber which allows a $\pi/K$ separation of about $1.5\sigma$.
The dominant physics backgrounds for the suppressed mode are the inclusive $B^- \to D^0 \pi^-$, with $D^0 \to X$ (where $X$ are modes other than $K\pi$); $B^- \to D^0 K^-$, with $D^0 \to X$; $B^- \to D^{0*} \pi^-$, with $D^{0*} \to D^0 \pi^0/ \gamma$; $B^- \to K^- \pi^+ \pi^-$ and $B^0 \to D^{*-}_0 l^+ \nu_l$.

\begin{figure}[!h]
%\begin{center}
\centering
\includegraphics[width=2.95in]{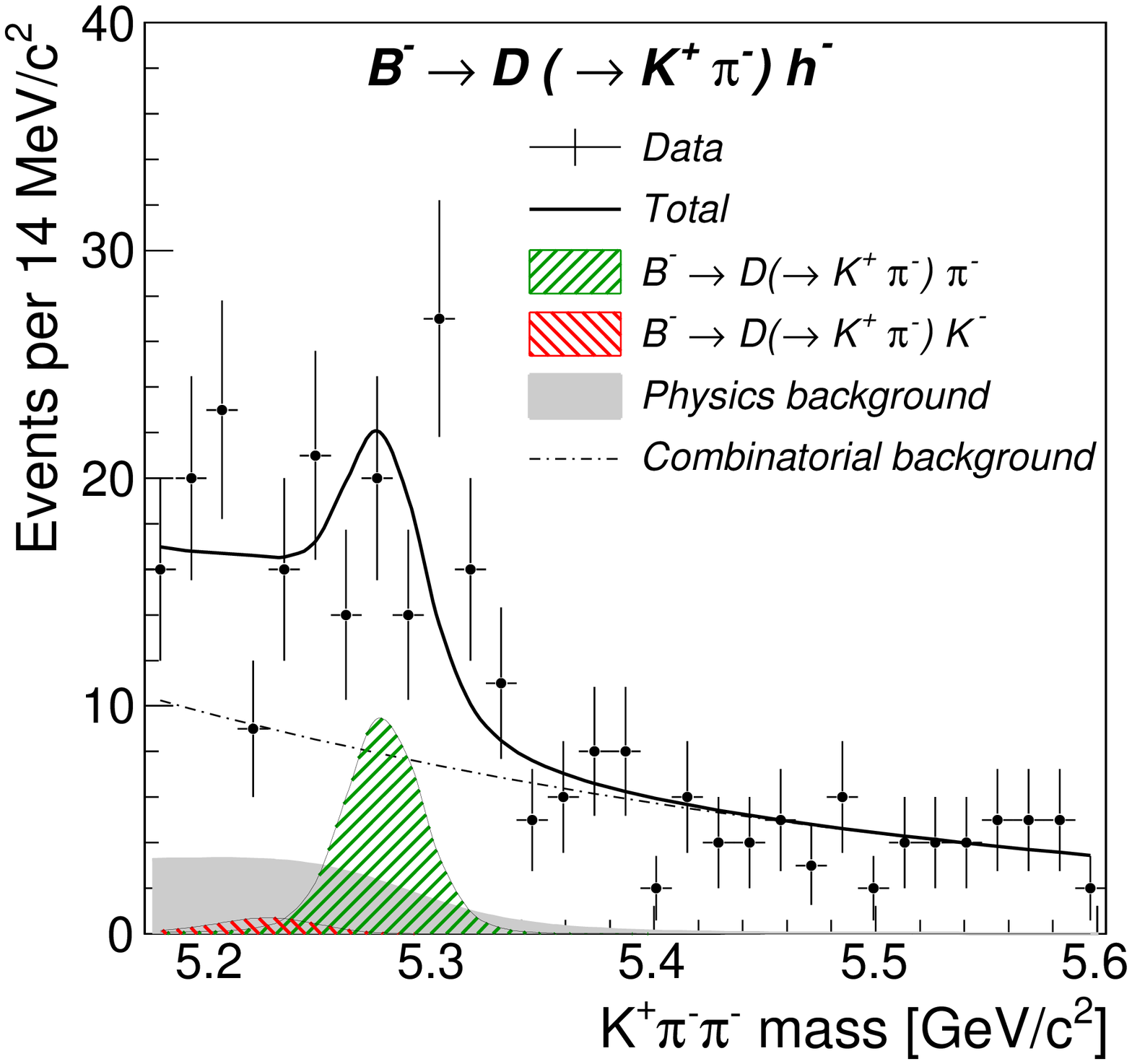} 
\hspace{0.5cm}
\includegraphics[width=2.95in]{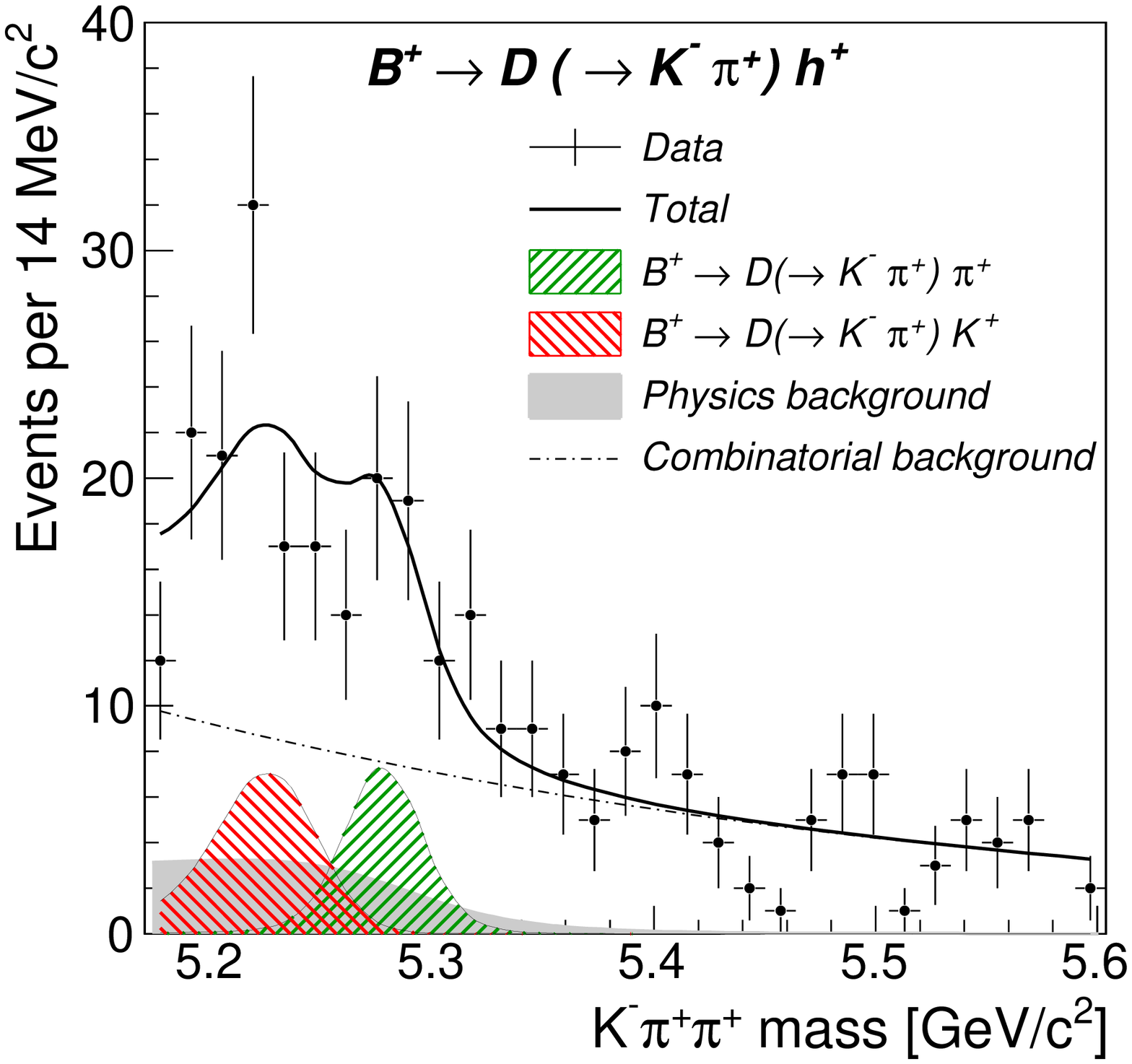} 
\caption{Mass distributions of $B^- \to D_{\it sup} h^-$ candidates for negative (left) and positive (right) charges. The projections of the likelihood fit are overlaid.} \label{fig:plot_dcs}
%\end{center}
\end{figure}

Projections of the fit in the suppressed mass distributions, separated in charge, are shown in Fig.~\ref{fig:plot_dcs}. The physics background contributions are summed in a single shape.
We obtained $1461 \pm 57$ $B^- \to D_{\it fav} K^- {\it (+ c.c.)}$, $19774 \pm 145$ $B^- \to D_{\it fav} \pi^- {\it (+ c.c.)}$,
$32 \pm 12$ $B^- \to D_{\it sup} K^- {\it (+ c.c.)}$ and $55 \pm 14$ $B^- \to D_{\it sup} \pi^- {\it (+ c.c.)}$ signal events. The significance of the $B^- \to D_{\it sup} K^-$ is 3.2 $\sigma$, including systematics, while the significance of the $B^- \to D_{\it sup} \pi^-$ signal is 3.6 $\sigma$.

Results on the observables have to be corrected for the different probabilities of $K^+$, $K^-$, $\pi^+$ and $\pi^-$ to interact with the tracker material.
We use previous measurements of $\frac{\epsilon(K^+)}{\epsilon(K^-)}=1.0178\pm 0.0023 (\rm{stat}) \pm 0.0045 (\rm{syst})$ and $ \frac{\epsilon(\pi^+)}{\epsilon(\pi^-)}=0.997\pm 0.003 (\rm{stat})\pm 0.006 (\rm{syst})$~\cite{ref:acp_corr}.
We extract $\frac{\epsilon (K^- \pi^+)}{\epsilon (K^+ \pi^-)} = 0.998 \pm 0.015(\rm{stat}) \pm 0.016(\rm{syst})$ from our own sample of favored $B^- \to D \pi^-$ decays.

The final results for the asymmetries are~\cite{ref:dcsPaper}
\begin{eqnarray}
A_{ADS}(K) & =  &  -0.82\pm 0.44(\rm{stat})\pm 0.09(\rm{syst}), \nonumber \\
A_{ADS}(\pi) & = &  0.13\pm 0.25(\rm{stat})\pm 0.02(\rm{syst}). \nonumber
\end{eqnarray}
The observed asymmetry of the kaon deviates from zero by 2.2 standard deviations.
\\
The ratios of suppressed to favored modes are 
\begin{eqnarray}
R_{ADS}(K)& = &[22.0\pm 8.6(\rm{stat})\pm 2.6(\rm{syst})]\times 10^{-3}, \nonumber \\
R^+ (K) &  = & [42.6\pm 13.7 (\rm{stat})\pm 2.8 (\rm{syst})]\times 10^{-3}, \nonumber \\
R^- (K)  & = & [3.8\pm 10.3 (\rm{stat})\pm 2.7 (\rm{syst})]\times 10^{-3}, \nonumber \\
R_{ADS}(\pi) & = & [2.8\pm 0.7(\rm{stat})\pm 0.4(\rm{syst})]\times 10^{-3},\nonumber \\
R^+ (\pi) &  = & [2.4\pm 1.0 (\rm{stat})\pm 0.4 (\rm{syst})]\times10^{-3}, \nonumber \\
R^- (\pi) &  = & [3.1\pm 1.1 (\rm{stat})\pm 0.4 (\rm{syst})]\times 10^{-3}. \nonumber 
\end{eqnarray}
These quantities are in agreement with existing measurements~\cite{ref:babarPaper,ref:bellePaper,ref:lhcbNote} and significantly contribute to the global knowledge of $\gamma$~\cite{ref:hfag,ref:ckmfitter}.

%\clearpage
%\newpage
%\newpage
\section{Gronau-London-Wyler method}
In the GLW method~\cite{ref:glw1,ref:glw2} the CP asymmetry of $B^- \to D_{CP\pm} K^-$ is studied, where $D$ is $D^0$ or $\overline{D}^0$ and $CP \pm$ are the $CP$ even and odd eigenstates of the $D$: $D_{CP^+} \rightarrow K^+ K^-, \pi^+ \pi^-$ and $D_{CP-} \rightarrow K^0_s \pi^0, K^0_s \phi, K^0_s \omega$.

We can define four observables:
\begin{eqnarray}
A_{CP \pm}   &=& \frac{\mathcal{B} (B^- \to D_{CP \pm} K^-) - \mathcal{B} (B^+ \to D_{CP \pm} K^+)} {\mathcal{B} (B^- \to D_{CP \pm} K^-) + \mathcal{B} (B^+ \to D_{CP \pm} K^+)}, \nonumber \\
 R_{CP \pm} &  = & 2 \cdot \frac{\mathcal{B} (B^- \to D_{CP \pm} K^-) + \mathcal{B} (B^+ \to D_{CP \pm} K^+)} {\mathcal{B} (B^- \to D_{\it fav} K^-) + \mathcal{B} (B^+ \to \overline{D}_{\it fav} K^+)}. \nonumber
 \end{eqnarray}
The relations with the amplitude ratios and phases are $A_{CP \pm}  =  2 r_B \sin{\delta_B} \sin{\gamma} / R_{CP \pm}$ and $R_{CP \pm} =  1 + r^2_B \pm   2 r_B \cos{\delta_B} \cos{\gamma} $.
Three of them are independent observables since $A_{CP+}R_{CP+} = -A_{CP-}R_{CP-}$.
%, to which correspond three unknowns. For this reason the method is very clean, 
Unfortunately the sensitivity to $\gamma$ is  proportional to $r_B$, so asymmetries are expected to be small.

\begin{figure}[!h]
%\begin{center}
\centering
%\includegraphics[width=2.9in]{figure/cf_glw.eps}
%\vspace{0.5cm}
\includegraphics[width=2.9in]{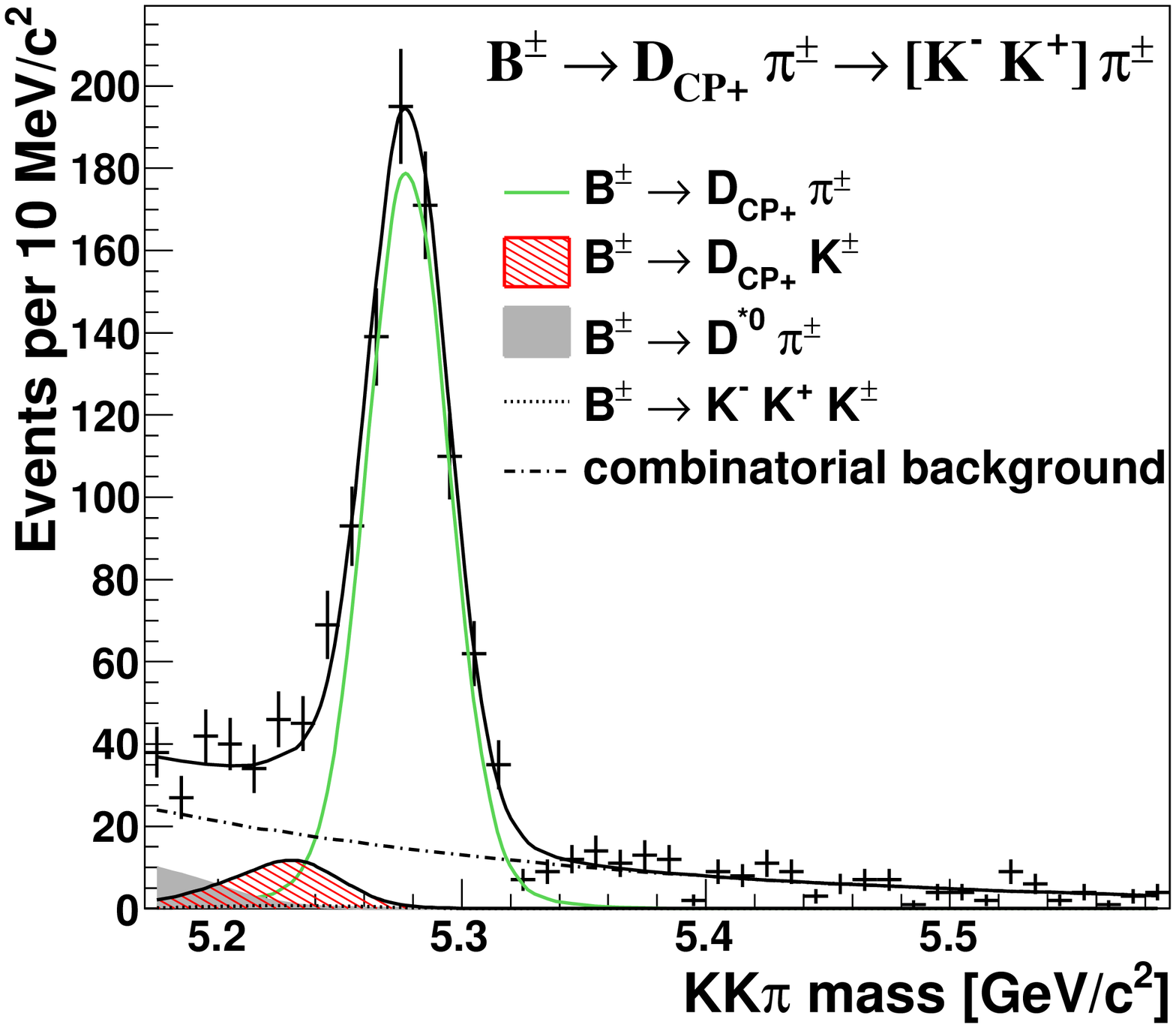}
\hspace{0.5cm}
\includegraphics[width=2.9in]{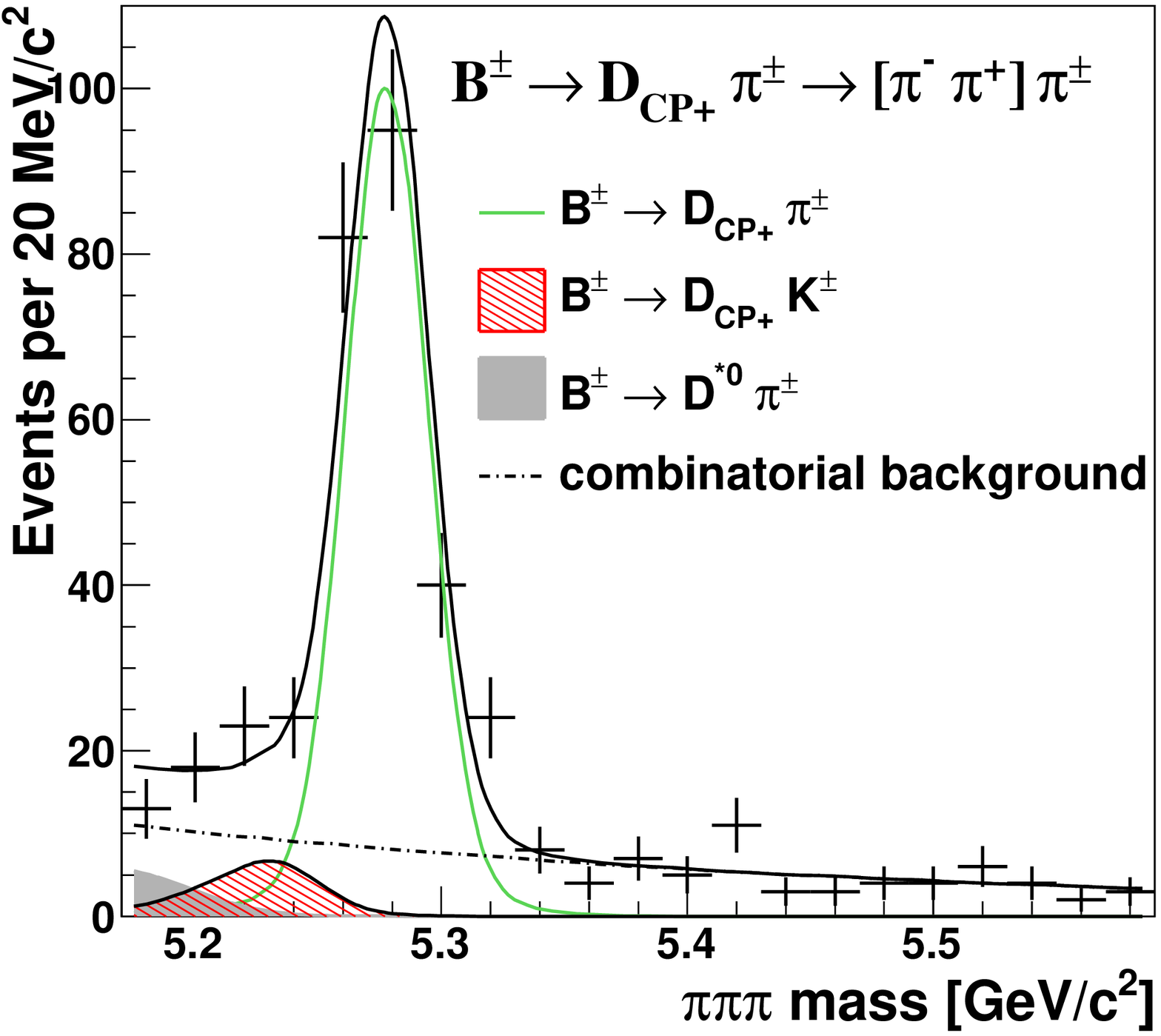} 
\caption{Mass distributions of $B^- \to D_{CP} h^-$ candidates for each reconstructed decay mode, Cabibbo-suppressed $K^+ K^-$ on the left and Cabibbo-suppressed $\pi^+ \pi^-$ on the right. The projections of the likelihood fit are overlaid for each mode.} \label{fig:plots_glw}
%\end{center}
\end{figure}

CDF performed the first measurement of branching fraction and $CP$ asymmetry of the $CP+$ modes at a hadron collider, using 1 fb$^{-1}$ of data~\cite{ref:dcp}.
The mass distributions obtained for the two modes of interest
($D \to K^+K^-$ and $\pi^+\pi^-$) are reported in Fig.~\ref{fig:plots_glw}, where a clear $B^- \to D \pi^-$ signal can be seen in each plot.

The dominant backgrounds are the combinatorial background
and the mis-reconstructed physics background such as $B^- \to D^{0*} \pi^-$ decay. In the
$D^0 \to K^+ K^-$ final state also the non-resonant $B^- \to K^-K^+K^-$ decay appears, as determined by a study
on CDF simulation~\cite{ref:notaDcp}. 
From an unbinned maximum likelihood fit, exploiting kinematic and particle
identification information, we obtained about 90 $B^- \to D_{CP +} K^-$ events and we measured the double ratio of CP-even to flavor eigenstate branching fractions and the direct CP asymmetry:
\begin{eqnarray}
R_{CP+} & = & 1.30 \pm 0.24 \mbox{(stat)} \pm  0.12 \mbox{(syst)}, \nonumber \\
A_{CP+} & = & 0.39 \pm 0.17 \mbox{(stat)} \pm  0.04 \mbox{(syst)}. \nonumber
\end{eqnarray}
These results are in agreement with previous measurements from $\Upsilon$(4S) decays~\cite{ref:hfag,ref:ckmfitter}.

\section{Conclusions}
The CDF experiment is pursuing a global program to measure the $\gamma$ angle from tree-dominated processes. 
The published measurement using the GLW method and the preliminary result using the ADS method show competitive results with previous measurements performed at $B$-factories and demonstrate the feasibility of these kinds of measurements also in a hadron collider environment.
 
We expect to increase the data-set available by the end of the year 2011 and obtain interesting and more competitive results in the near future.

\end{document}